**Multi-dimensional band structure spectroscopy in the synthetic frequency dimension**

Dali Cheng,[1] Eran Lustig,[1] Kai Wang,[1,2] and Shanhui Fan[1,*]

[1]*Ginzton Laboratory and Department of Electrical Engineering, Stanford University, Stanford, California 94305, USA*

[2]*Department of Physics, McGill University, Montreal, QC H3A 2T8, Canada*

[*]shanhui@stanford.edu

**Abstract**

The concept of synthetic dimensions in photonics provides a versatile platform in exploring multi-dimensional physics. Many of these physics are characterized by band structures in more than one dimensions. Existing efforts on band structure measurements in the photonic synthetic frequency dimension however are limited to either one-dimensional Brillouin zones or one-dimensional subsets of multi-dimensional Brillouin zones. Here we theoretically propose and experimentally demonstrate a method to fully measure multi-dimensional band structures in the synthetic frequency dimension. We use a single photonic resonator under dynamical modulation to create a multi-dimensional synthetic frequency lattice. We show that the band structure of such a lattice over the entire multi-dimensional Brillouin zone can be measured by introducing a gauge potential into the lattice Hamiltonian. Using this method, we perform experimental measurements of two-dimensional band structures of a Hermitian and a non-Hermitian Hamiltonian. The measurements reveal some of the general properties of point-gap topology of the non-Hermitian Hamiltonian in more than one dimensions. Our results demonstrate experimental capabilities to fully characterize high-dimensional physical phenomena in the photonic synthetic frequency dimension.

## 1. Introduction

In the synthetic dimensions in photonics [1-20], different internal degrees of freedom of photons are coupled to form extra dimensions in addition to the dimensions of the real space. Using this concept, one can experimentally study novel physical phenomena unique to high-dimensional systems with low-dimensional platforms, which are of less complexity in engineering and control. Recent experimental accomplishments in photonic synthetic dimensions include the demonstration of photonic analogues of quantum Hall effect and topological insulators [5,8,9,11,16], as well as the realization of the skin effect and eigenvalue topologies in non-Hermitian systems [21-23].

In the experimental demonstration of many of these physical phenomena, band structure measurements are essential since much of the nontrivial physics of the systems manifests in the band structure [11,22-26]. In the synthetic frequency dimension [7], current experimental band structure measurements are carried out in either the one-dimensional Brillouin zone [11,22-25,27,28] or a one-dimensional subset of the two- or three-dimensional Brillouin zone [29]. In this paper, we propose and demonstrate the method of multi-dimensional band structure spectroscopy in the synthetic frequency dimension. We demonstrate that the band structure over the *entire* multi-dimensional Brillouin zone can be probed, by exploiting a modulation phase which introduces a gauge potential in the Hamiltonian. As examples, we measure the two-dimensional band structures of both a Hermitian Hamiltonian and a non-Hermitian Hamiltonian. Our measurement reveals some of the general properties of point-gap topology of the non-Hermitian Hamiltonian in more than one dimensions. The results here demonstrate experimental capabilities to fully characterize high-dimensional physical phenomena in the photonic synthetic frequency dimension.

## 2. Theory of multi-dimensional Brillouin zone sampling

We here describe the theory of multi-dimensional band structure measurements in the synthetic frequency dimension. As an illustration we consider a Hamiltonian that describes a two-dimensional square lattice:

$$H = \frac{1}{2} \sum_{x,y} \left( g a_{x+1,y}^\dagger a_{x,y} + \kappa a_{x,y+1}^\dagger a_{x,y} + \text{H.c.} \right), \qquad x, y \in \mathbb{Z}$$

(1)

Here, $a_{x,y}^\dagger$ and $a_{x,y}$ are creation and annihilation operators on the $(x,y)$ lattice site, respectively. $g$ and $\kappa$ are real coupling constants along the $x$ and $y$ directions. The corresponding band structure is

$$E(k_x, k_y) = g \cos k_x + \kappa \cos k_y$$

(2)

Strictly speaking, the band structure of Eq. (2) is applicable only when the lattice is of infinite size along both $x$ and $y$ directions, in which case the wavevectors $k_x$ and $k_y$ each occupies the entire interval of $(-\pi, \pi]$.

Our objective is to experimentally create and fully characterize multi-dimensional band structures such as Eq. (2) using the approach of synthetic frequency dimensions. We first review the approach to create and measure a one-dimensional band structure in the synthetic frequency dimension. As shown in Fig. 1(a), one uses a photonic ring resonator under dynamical modulation. The resonator by itself, without the modulation, supports a set of longitudinal modes that are equally spaced by the free spectral range $\Omega_R = 2\pi/T_R$, where $T_R$ is the round-trip time of light propagation inside the resonator. Here we assume that both polarizations are degenerate in the resonator, that the group velocity dispersion is absent, and that the sizes of the modulators are negligible compared to the circumference of the resonator. When the resonator is modulated at $\Omega_R$, resonator modes separated in frequency by $\Omega_R$ are coupled. Taking each mode as a lattice site, the dynamics of the mode amplitudes can be modelled by a one-dimensional tight-binding lattice along the synthetic frequency dimension [7]. To probe the band structure of this one-dimensional lattice, the resonator is excited by a continuous wave (CW) laser input with tunable frequency $\omega_{CW} = \omega_0 + n\Omega_R + \delta\omega$, through an input-output waveguide that couples to the ring with power coupling ratio $\gamma$. Here $\omega_0$ is the central frequency, $n \in \mathbb{Z}$ and $\delta\omega$ is the frequency detuning. For each detuning $\delta\omega$, the light intensity $\xi$ at an output port can be measured as a function of $t$, where $t \in (-T_R/2, T_R/2]$ is the time variable within each round-trip. By interpreting the time variable $t$ as the wavevector $k$ along the frequency axis, the band structure of this one-dimensional lattice can be extracted from the resonant features in the output intensity $\xi(\delta\omega, t)$ [24].

Yuan *et al*. theoretically proposed that multi-dimensional Hamiltonians can also be synthesized using a single resonator as shown in Fig. 1(a) [30]. To implement the Hamiltonian in Eq. (1), Yuan *et al*. considered a phase modulator with the transmission coefficient:

$$\tau_{\text{ref}}(t) = e^{i[gT_R \cos(\Omega_R t) + \kappa T_R \cos(M\Omega_R t)]}$$

(3)

Here $M > 1$ is an integer. Since the modulation waveform contains both $\Omega_R$ and $M\Omega_R$ frequency components, the $m^{\text{th}}$ frequency mode in the resonator is coupled to both $(m \pm 1)^{\text{th}}$ mode and $(m \pm M)^{\text{th}}$ mode. The corresponding picture of a one-dimensional lattice in the synthetic frequency dimension is shown in Fig. 1(b) on the left, with $M = 5$. The 1$^{\text{st}}$-order and $M^{\text{th}}$-order couplings are shown in brown and red, respectively. In this lattice, the $M^{\text{th}}$-order coupling can be viewed as the hopping along an additional synthetic frequency dimension. By rearranging the positions of the lattice sites, one can see that the one-dimensional lattice in Fig. 1(b) on the left is equivalent to a two-dimensional square lattice with nearest-neighbor coupling in Fig. 1(b) on the right. This square lattice is infinite along the $y$ direction and has a finite size of $M$ lattice sites along the $x$ direction with a twisted boundary condition imposed on the edges [14,30]. In this way, one can synthesize multi-dimensional lattices in a single resonator by additional modulation frequencies, although such lattices are in nature of finite size along all synthetic frequency dimensions except one.

In such multi-dimensional synthetic lattices with a twisted boundary condition as shown in Fig. 1(b), the allowed wavevectors form a one-dimensional subset of the two-dimensional Brillouin zone of the corresponding infinite two-dimensional lattice. Consider an eigenstate $\psi(x, y)$ in the two-dimensional lattice in Fig. 1(b) on the right, with $1 \leq x \leq M$ and $x, y \in \mathbb{Z}$. We define the translation operators along $x$ and $y$ directions:

$$\hat{T}_x \psi(x, y) = \begin{cases} \psi(x+1, y), & x \neq M \\ \psi(1, y+1), & x = M \end{cases}$$

$$\hat{T}_y \psi(x,y) = \psi(x, y+1)$$

(4)

According to Bloch's Theorem, we can write $\psi(x, y+1)$ in terms of $\psi(x,y)$ in two alternative ways: $\psi(x, y+1) = \hat{T}_y \psi(x,y) = e^{ik_y}\psi(x,y)$, and $\psi(x, y+1) = \hat{T}_x^M \psi(x,y) = e^{iMk_x}\psi(x,y)$. Here $k_x$ and $k_y$ are wavevectors along the $x$ and $y$ directions of the corresponding infinite two-dimensional lattice. Therefore $k_y = Mk_x$ (mod $2\pi$). In such a two-dimensional synthetic lattice under twisted boundary condition, the allowed wavevectors sample the first Brillouin zone of the corresponding infinite two-dimensional lattice at a set of line segments:

$$S_0 = \{(k_x, k_y) \mid k_y = Mk_x \text{ (mod } 2\pi), -\pi < k_x, k_y \leq \pi\}$$

(5)

We notice that the multi-dimensional lattices proposed in [30] has been realized experimentally in [29] with a large $M$ on the order of 100. Ref. [29] provided results on the allowed wavevectors (dashed lines in Fig. 1(c), $S_{\text{ref}}$) that differ from Eq. (5) (solid lines in Fig. 1(c), $S_0$), but the differences vanish in the large-$M$ limit. Our result here is applicable for any value of $M$.

As can be seen from Eq. (5) above, a key limitation of the existing scheme for multi-dimensional band structure measurements is that the high dimensional Brillouin zone is not fully probed. Here we show that this limitation can be overcome by introducing a reconfigurable gauge potential into the Hamiltonian. As an illustration, to fully measure the band structure of Eq. (2) over the entire two-dimensional Brillouin zone, instead of Eq. (3), we set the transmission coefficient of the phase modulator as

$$\tau(t) = e^{iV(t)} = e^{i[gT_R \cos(\Omega_R t) + \kappa T_R \cos(M\Omega_R t + \varphi)]}$$

(6)

We show that the entire two-dimensional Brillouin zone can be fully sampled by varying both the time variable $t$ and the modulation phase $\varphi$ which operates as a gauge potential in the Hamiltonian [31]. In the weak-coupling limit $\gamma \ll 1$, the steady-state transmission function can be expressed as [23]

$$\xi(\delta\omega, t) \approx 1 - 2\gamma \text{Re}\left\{\frac{1}{e^{\gamma_0} e^{i[\delta\omega T_R - V(t)]} - 1}\right\}$$

(7)

where $\gamma_0 > 0$ represents the round-trip intrinsic loss of the resonator. The transmission function reaches minima when $\delta\omega T_R - V(t) = 0$, i.e.,

$$\delta\omega = g \cos(\Omega_R t) + \kappa \cos(M\Omega_R t + \varphi)$$

(8)

By comparing Eq. (2) and Eq. (8), we see that the locations of the minima of the transmission function correspond to the band energies $E(k_x, k_y)$, if we make the substitution

$$k_x = \Omega_R t, \quad k_y = M\Omega_R t + \varphi \text{ (mod } 2\pi)$$

(9)

By measuring the transmission function $\xi(\delta\omega, t)$, for a given $\varphi$, one probes the band structure at a set of line segments:

$$S_\varphi = \{(k_x, k_y) \mid k_y = Mk_x + \varphi \pmod{2\pi}, -\pi < k_x, k_y \leq \pi\} \tag{10}$$

Fig. 2(a) shows in the Brillouin zone the sets $S_\varphi$ with $M = 5$ and $\varphi = 0, \pi$. Note that $S_0$ is a special case of $S_\varphi$ with $\varphi = 0$, as the gauge potential $\varphi$ was not introduced in the derivation of Eq. (5). Since the modulation phase $\varphi$ is a continuous variable that can be externally controlled, by properly choosing the values of $\varphi$, we can reconstruct the band energy at every point in the Brillouin zone:

$$\bigcup_{\varphi \in [0, 2\pi)} S_\varphi = (-\pi, \pi] \otimes (-\pi, \pi] \tag{11}$$

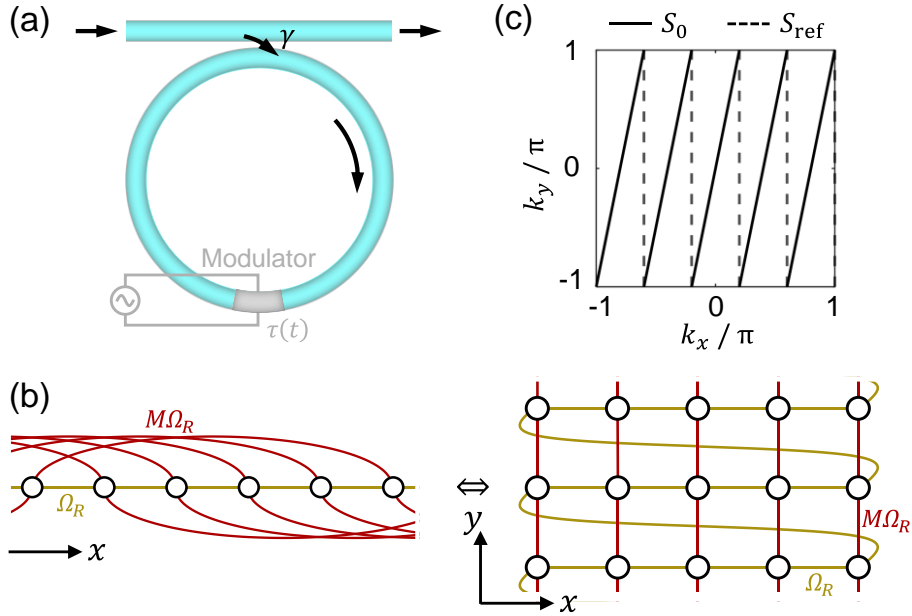

**Fig. 1.** Concept of the multi-dimensional synthetic frequency lattice and Brillouin zone sampling. (a) The experimental platform of a photonic resonator under dynamical modulation. $\tau(t)$ is the transmission coefficient of the modulation. (b) The tight-binding lattice with both $1^{st}$- and $M^{th}$-order coupling in the one-dimensional synthetic frequency space, and the equivalent square lattice with nearest-neighbor coupling and twisted boundary condition in the two-dimensional synthetic frequency space. (c) The sets of line segments $S_0$ and $S_{\text{ref}}$ [29] in the two-dimensional Brillouin zone at which the wavevectors are sampled.

## 3. Experimental measurement of a two-dimensional Hermitian band structure

Based on the theory above, here we provide an experimental demonstration of measurement of a two-dimensional Hermitian band structure in the synthetic frequency dimension. Our setup is similar to those in [22-24]. More detailed discussions can be found in section 5. The resonator is an optical fiber cavity with the free spectral range $\Omega_R = 2\pi \times 6$ MHz. The resonator is excited by a narrow-linewidth continuous wave laser through a fiber coupler, and the laser frequency can be tuned within a range much larger than $\Omega_R$. Inside the resonator, we incorporate lithium-niobate-based electro-optic modulators (EOMs), and an erbium-doped fiber amplifier (EDFA) to partially compensate for the cavity loss. The light intensity $\xi$ at the transmission port is measured by a large-bandwidth photodiode as a function of laser frequency and time.

Using the experimental setup, we perform band structure measurements on the two-dimensional lattice in Fig. 1(b). Here we take $gT_R = 2\kappa T_R = 0.12$, and the band structure of our interest is $E_1(k_x, k_y) = 2\kappa \cos k_x + \kappa \cos k_y$. Based on the theory in the previous section, to synthesize and measure this band structure, we choose the transmission coefficient of the phase modulator as

$$\tau_1(t) = e^{i\kappa T_R[2\cos(\Omega_R t) + \cos(M\Omega_R t + \varphi)]}$$

(12)

Given a specific value of $\varphi$, the measured transmission function $\xi(\delta\omega, t)$ reveals the band energies sampled at $S_\varphi$. By taking multiple measurements with different values of $\varphi$, we can characterize the two-dimensional band structure $E_1(k_x, k_y)$ over the entire first Brillouin zone in two dimensions.

The right panels in Figs. 2(b) and 2(c) show the transmission function $\xi(\delta\omega, t)$ at $\varphi = 0$ and $\varphi = \pi$, respectively. The left panels show the corresponding line plots for $\xi(\delta\omega, t = 0)$. At each $t$, the transmission function exhibits a periodic set of resonant features, and the frequency difference between the neighboring transmission minima is $\Omega_R$. The periodic behavior here arises from the translational symmetry of the lattice in Fig. 1(b) along the synthetic frequency dimension. Fig. 2(d) shows the frequency locations of the transmission minima as a function of $t$ for $\varphi = 0$ and $\varphi = \pi$. We see that these locations shift as $\varphi$ varies. These locations, as mentioned above, correspond to the band energy $E_1(k_x, k_y)$ sampled at $S_\varphi$. In Fig. 2(e), we reconstruct the measured band structure in the two-dimensional Brillouin zone according to Eq. (10). The measured band structure agrees well with the theoretical value represented by the gray surface.

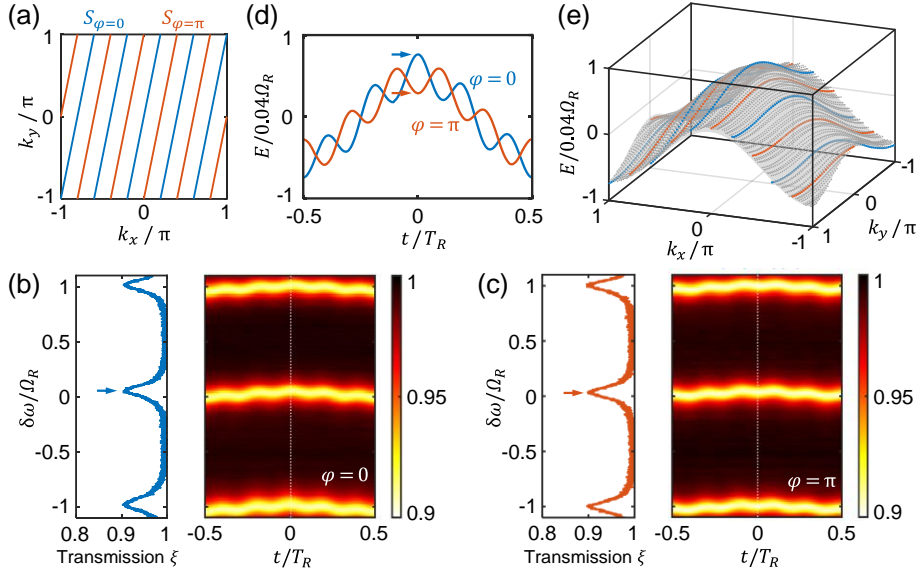

**Fig. 2.** Two-dimensional band structure measurement of the Hermitian Hamiltonian $H_1$. (a) The set of line segments $S_\varphi$ in the two-dimensional Brillouin zone. (b),(c) Measured transmission functions $\xi(\delta\omega, t)$ at (b) $\varphi = 0$ and (c) $\varphi = \pi$. The sliced spectra at $t = 0$, indicated by the white dashed lines in the colormap plots, are shown on the left. (d) The fitted frequency locations of the transmission minima (corresponding to band energies) as a function of $t$. (e) The reconstructed band structure. Blue and orange dots correspond to the modulation with $\varphi = 0$ and $\varphi = \pi$, respectively, and gray dots correspond to other values of $\varphi$. Here we take $\varphi \in \{0, 0.1\pi, 0.2\pi, \cdots, 1.8\pi, 1.9\pi\}$. The gray surface represents the theoretical band structure $E_1(k_x, k_y)$. $M = 5$ in this figure.

## 4. Experimental measurement of a two-dimensional non-Hermitian band structure and its point-gap topology

We now use the capability, as demonstrated above, to explore novel topological features of non-Hermitian band structures in two dimensions. A key novel aspect of non-Hermitian band structure is that the eigenvalues, being complex, exhibit nontrivial topology in the wavevector space [32-35]. In one-dimension, non-Hermitian systems can exhibit point-gap topology where the eigenvalues form nontrivial contours as the wavevector varies across the first Brillouin zone. This point-gap topology was first experimentally demonstrated in the synthetic frequency dimension in [22], and is related to the non-Hermitian skin effect [34,36-38] when the lattice is truncated due to the bulk-boundary correspondence. It was also noted theoretically that such point-gap topology can also exist in higher dimensional systems [39-43], with additional constraints due to the geometry of the Brillouin zone. Here we provide an experimental demonstration of point-gap topology in two-dimensional systems which has not been previously carried out.

We consider the non-Hermitian Hamiltonian

$$H_2 = \frac{\mu + i\eta}{2} \sum_{x,y} (a^\dagger_{x+1,y} a_{x,y} + a^\dagger_{x-1,y} a_{x,y}) + \frac{\mu - i\eta}{2} \sum_{x,y} (a^\dagger_{x,y+1} a_{x,y} + a^\dagger_{x,y-1} a_{x,y}), \qquad x, y \in \mathbb{Z}$$

where $\mu, \eta$ are real coupling constants. This Hamiltonian is described by a complex symmetric matrix and hence is reciprocal. The band structure of this Hamiltonian is

$$E_2(k_x, k_y) = \mu(\cos k_x + \cos k_y) + i\eta(\cos k_x - \cos k_y) \tag{14}$$

Such band structure can exhibit nontrivial point-gap topology. In the two-dimensional Brillouin zone, we define a set of line segments

$$L(\alpha, \delta k) = \{(k_x, k_y) \mid k_y = \alpha k_x + \delta k \pmod{2\pi}, \ \alpha \in \mathbb{Q} \cup \{\infty\}, \ \delta k \in \mathbb{R}, \ -\pi < k_x, k_y \leq \pi\} \tag{15}$$

where $\alpha$ is the direction (slope) of the line segments and is assumed to be a rational number. These line segments are connected at the edges of the Brillouin zone, and topologically, $L(\alpha, \delta k)$ forms a closed loop in the two-dimensional Brillouin zone $\mathbb{T}^2$. Associated with $L(\alpha, \delta k)$ we can define the winding number [33,41]

$$w(L(\alpha, \delta k), E_0) \equiv \frac{1}{2\pi i} \oint_{L(\alpha, \delta k)} dk \frac{\partial}{\partial k} \log[E_2(k) - E_0] \tag{16}$$

where $E_0 \in \mathbb{C}$ is the reference energy. A non-zero winding number indicates a nontrivial point-gap topology along the loop.

As was noted in [41,42], in high dimensions there are general theoretical results about these winding numbers, depending on the nature of the loops. For a reciprocal Hamiltonian, its band structure satisfies $E(-k_x, -k_y) = E(k_x, k_y)$, and therefore

$$w(L, E_0) = -w(\tilde{L}, E_0) \tag{17}$$

where $\tilde{L}$ is the time-reversal partner of $L$. And for a time-reversal invariant loop with $L = \tilde{L}$, the winding number is zero. The aim of our experiments in this section is to demonstrate these theoretical observations.

To implement the non-Hermitian band structure $E_2(k_x, k_y)$, we incorporate both phase and amplitude modulations in the resonator. The transmission coefficients of the modulators are

$$\tau_2^{\text{phs}}(t) = e^{i\mu T_R[\cos(\Omega_R t) + \cos(M\Omega_R t + \varphi)]}, \qquad \tau_2^{\text{amp}}(t) = e^{-\eta T_R[\cos(\Omega_R t) - \cos(M\Omega_R t + \varphi)]} \tag{18}$$

where $\mu T_R = 0.09$ and $\eta T_R = 0.08$. With different values of $\varphi$, we take multiple measurements of the transmission function $\xi(\delta\omega, t)$ to reconstruct $E_2(k_x, k_y)$. Note that the imaginary part of the band energy is associated with the linewidth of the resonant features in the spectrum $\xi(\delta\omega, t)$ for a given $t$ [22].

Figure 3 presents the measured band structure with the modulation Eq. (18). Fig. 3(a) shows the transmission function with $\varphi = \pi$, with $\xi(\delta\omega, t = -0.45T_R)$ and $\xi(\delta\omega, t = 0.05T_R)$ displayed on the left as examples. With dynamic amplitude modulation, the instantaneous loss rate inside the resonator is not a constant, and in this example the spectrum $\xi(\delta\omega, t = -0.45T_R)$ has a sharper line shape than $\xi(\delta\omega, t = 0.05T_R)$. By extracting the locations and linewidths of the resonant features as a function of $t$, we obtain $\text{Re}(E_2)$ and $\text{Im}(E_2)$ along the set of line segments $S_\varphi$ in the two-dimensional Brillouin zone. In Fig. 3(b), we plot the real and imaginary parts of the reconstructed band structure $E_2(k_x, k_y)$, in agreement with the theoretical values represented by the gray surfaces.

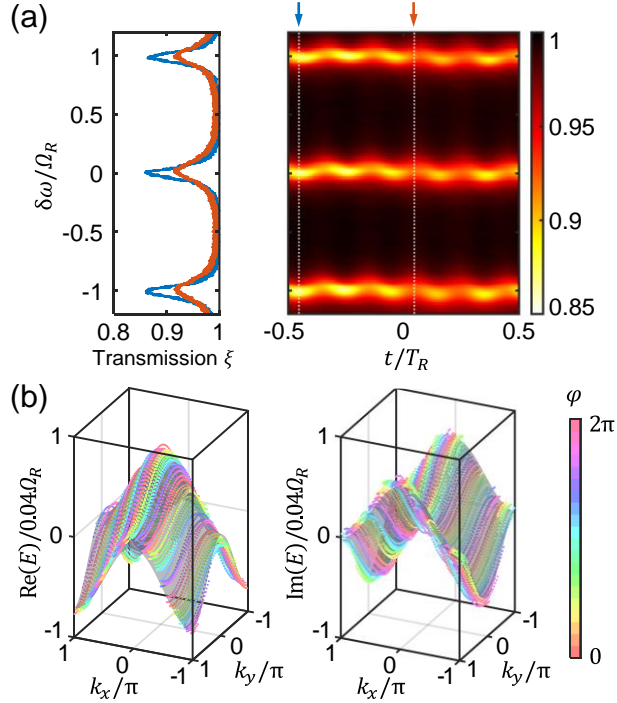

**Fig. 3.** Two-dimensional band structure measurement of the non-Hermitian Hamiltonian $H_2$. (a) Measured transmission function $\xi(\delta\omega, t)$ at $\varphi = \pi$. The sliced spectra at $t = -0.45T_R$ and $t = 0.05T_R$, indicated by the white dashed lines, are shown on the left in blue and orange, respectively. (b) The reconstructed real and imaginary parts of the band structure. Here we take $\varphi \in \{0, 0.1\pi, 0.2\pi, \cdots, 1.8\pi, 1.9\pi\}$, and dots of different colors correspond to different values of $\varphi$. The gray surfaces represent the theoretical band structure $\text{Re}\{E_2(k_x, k_y)\}$ and $\text{Im}\{E_2(k_x, k_y)\}$. $M = 5$ in this figure.

To demonstrate Eq. (17), we choose the loop $L$ described by $E_0 = 0$ and $\alpha = 1$ in Eq. (16). Here we write $w(L(\alpha = 1, \delta k), E_0 = 0)$ as $w(\delta k)$ for convenience. From Eq. (17) we have

$$w(-\delta k) = -w(\delta k)$$

(19)

And for time-reversal invariant loops, i.e., $\delta k = 0$ or $\delta k = \pi$, the winding number is zero. In Fig. 4, we demonstrate the winding property stated by Eq. (19). For a given $\delta k$, we extract from Fig. 3(b) the complex energy at each point along the loop $L(\alpha = 1, \delta k)$, and the trajectory of the complex energies form the winding diagram. In Fig. 4, we notice that when $\delta k = 0$ or $\delta k = \pi$, the complex energies lie on a line and thus the winding number is indeed zero. For other values of $\delta k$, the winding is nontrivial. In particular, for $\delta k = -\pi/2$ and $\delta k = \pi/2$, the windings are of the same geometric shape but in opposite directions. This agrees with Eq. (19). In fact, for this Hamiltonian $H_2$, we always have $w(\delta k) = 1$ when $0 < \delta k < \pi$, and $w(\delta k) = -1$ when $-\pi < \delta k < 0$.

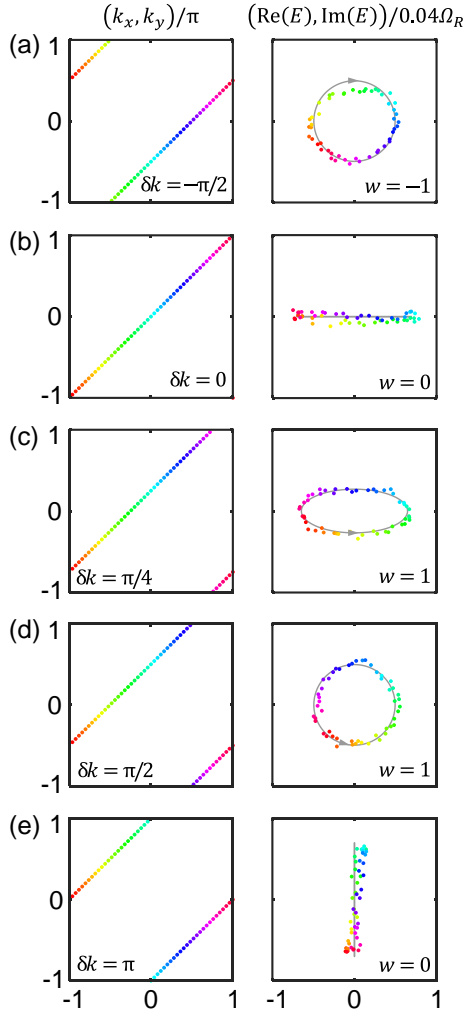

**Fig. 4.** Observation of the properties of point-gap topology of the non-Hermitian Hamiltonian $H_2$. The left column illustrates $L(\alpha = 1, \delta k)$ where the winding numbers are defined. The right column shows the winding diagrams. The grey solid lines in the right column are theoretical results. A complex energy on the right is related to its location in the Brillouin zone on the left by the color. (a) $\delta k = -\pi/2$, (b) $\delta k = 0$, (c) $\delta k = \pi/4$, (d) $\delta k = \pi/2$, (e) $\delta k = \pi$.

## 5. Methods

*Experimental setup.*—The schematic of our experimental platform is shown in Fig. 5. The setup is based on optical fibers. The resonator has a free spectral range of $\Omega_R = 2\pi \times 6$ MHz. Our light source is a grade 3 Orion laser from Redfern integrated optics, a continuous wave laser in the telecommunication C-band with a center wavelength of 1542.057 nm and a linewidth of 2.8 kHz. The frequency of the laser is tunable and controlled by a function generator, which generates a ramp voltage signal of 600 mV$_{pp}$ amplitude and 100 Hz frequency. The laser frequency is thus swept in a range of approximately $13\Omega_R$. The coupling between the input-output waveguide and the resonator is implemented by a $2 \times 2$ fiber coupler (beam splitter) of 95:5 power coupling ratio. Inside the resonator, light goes through phase and amplitude modulators, a polarization controller, an erbium-doped fiber amplifier (EDFA), and a dense wavelength-division multiplexing (DWDM) band-pass filter. The modulators are based on the electro-optic modulation in lithium niobate waveguides. The modulation signals are generated by Red Pitaya STEMlab field-programmable gate arrays (FPGAs) with 60 mV$_{pp}$ amplitude, and then amplified by Mini-Circuits ZHL-3A+ coaxial radiofrequency (RF) amplifiers and applied to the modulators. The polarization controller ensures that the light polarization remains unchanged after one round-trip propagation inside the resonator. The EDFA is used to partially compensate for the intrinsic loss in the resonator. We use a lower gain in the experiment to avoid gain saturation of the EDFA and lasing of the cavity. The band-pass filter, in channel 44, has a center wavelength of 1542.14 nm and a bandwidth of 26.5 GHz. It is used to suppress the amplified spontaneous emission noise of the EDFA, and supports approximately $4.4 \times 10^3$ frequency modes in its transmission bandwidth. Finally, light at the transmission port is pre-amplified by a semiconductor optical amplifier (SOA) to improve the signal-to-noise ratio, and then detected by a photodiode (PD) with 5 GHz bandwidth. The detected signal is collected by an oscilloscope with a sampling rate of 2 GSa/s.

*Data processing.*—The transmission function $\xi(\delta\omega, t)$ is dependent on both the laser detuning $\delta\omega$ and the time variable $t$. By making $\delta\omega$ linearly dependent on time, we can obtain the entire $\xi(\delta\omega, t)$ function within a single-shot measurement [24] where we sweep the laser frequency by a ramp signal. When processing the detected signal from the oscilloscope, we sequentially truncate the temporal signal into time intervals of length $T_R$. We assume that the laser frequency is unchanged within each time interval, but is different for different time intervals. Such approximation is valid since the laser frequency only changes by $2 \times 10^{-4} \Omega_R$ between neighboring time intervals. In Figs. 2 and 3, for each $\varphi$ we take a measurement of $\xi(\delta\omega, t)$. Given a time $t$, we use the Lorentzian line shape to fit the spectrum $\xi$ as a function of $\delta\omega$, and obtain the locations of the transmission minima and the linewidth [22]. The intrinsic loss $\gamma_0$ of the resonator contributes to the fitted linewidth, and is removed as a background constant when calculating the imaginary part of the band structure. We also take into consideration the delay times that the RF signals propagate from the output ports of the FPGA to the input ports of the modulators, and the frequency-dependent phase responses of the RF amplifiers.

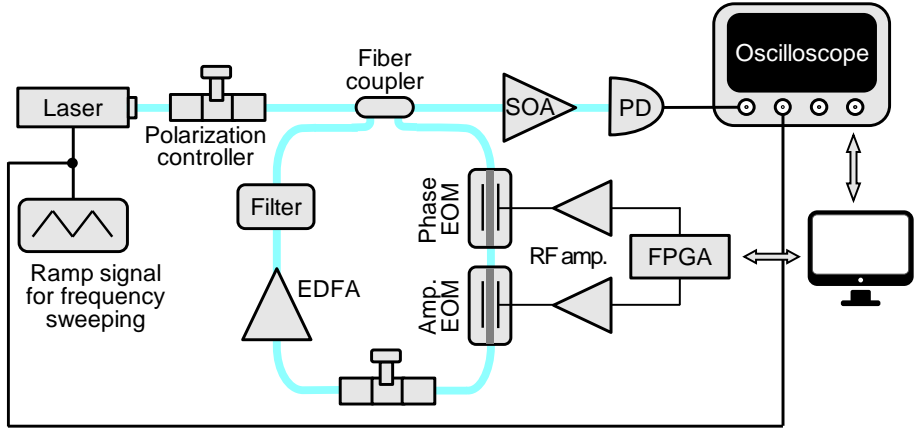

**Fig. 5.** Schematic of the experimental setup. The optical fibers are shown in blue. To create a Hermitian band structure, only the phase modulator is used, and to create a non-Hermitian band structure, both phase and amplitude modulators are used. EOM, electro-optic modulator; EDFA, erbium-doped fiber amplifier; SOA, semiconductor optical amplifier; PD, photodiode; FPGA, field-programmable gate array; RF amp., radiofrequency amplifier.

**6. Conclusions**

In this paper, we have proposed and demonstrated the method of multi-dimensional band structure spectroscopy in the photonic synthetic frequency dimension. By varying the modulation phase, we can fully reconstruct the band structure at each point in the multi-dimensional Brillouin zone. As examples, we create and fully characterize the band structures of both a Hermitian and a non-Hermitian tight-binding Hamiltonian on two-dimensional square lattices. We also observe some of the general properties of point-gap topology of the non-Hermitian Hamiltonian in more than one dimensions. Our method can also be applied to models of more complexity, for example, models with multiple bands in higher dimensions and with more complicated connectivity between the lattice sites.

**Acknowledgements**

We thank Professor David A. B. Miller for providing laboratory space and equipment. This work is supported by MURI projects from the U.S. Air Force Office of Scientific Research (Grants No. FA9550-18-1-0379 and FA9550-22-1-0339).